\documentclass[a4paper,twoside,prd,showpacs,nofootinbib,preprintnumbers,twocolumn]{revtex4-1}
\usepackage{amssymb,amsmath,bm,natbib}
\usepackage{color}
\usepackage{slashed}
\usepackage{graphics}
\usepackage{graphicx}
\usepackage[utf8]{inputenc}
\usepackage[caption=false]{subfig}
\usepackage{hyperref}
\usepackage{url}
\usepackage{dsfont}
\usepackage{float}
\usepackage{cancel}
\usepackage{xcolor,colortbl}
\usepackage{rotating}
\usepackage[export]{adjustbox}
\usepackage{flushend}

\newcommand{\eq}[1]{Eq.~\eqref{#1}}

\newcommand{\e}{\ensuremath{\mathrm{e}}}

\definecolor{Gray}{gray}{0.7}

\begin{document}
\preprint{PSI-PR-19-26,   UZ-TH  53/19}
	\title{Global Fit to Modified Neutrino Couplings and the Cabibbo-Angle Anomaly}

\author{Antonio M. Coutinho}
\email{antonio.coutinho@psi.ch}
\affiliation{Paul Scherrer Institut, CH--5232 Villigen PSI, Switzerland}

\author{Andreas Crivellin}
\email{andreas.crivellin@cern.ch}
\affiliation{Paul Scherrer Institut, CH--5232 Villigen PSI, Switzerland}
\affiliation{Physik-Institut, Universit\"at Z\"urich, Winterthurerstrasse 190, CH--8057 Z\"urich, Switzerland}
		
\author{Claudio Andrea Manzari}
\email{claudioandrea.manzari@physik.uzh.ch}
\affiliation{Paul Scherrer Institut, CH--5232 Villigen PSI, Switzerland}
\affiliation{Physik-Institut, Universit\"at Z\"urich, Winterthurerstrasse 190, CH--8057 Z\"urich, Switzerland}

\begin{abstract}

Recently, discrepancies of up to $4\,\sigma$ between the different determinations of the Cabibbo angle were observed. In this context, we point out that this ``Cabibbo-angle anomaly'' can be explained by lepton flavour universality violating new physics in the neutrino sector. However, modified neutrino couplings to standard model gauge bosons also affect many other observables sensitive to lepton flavour universality violation, which have to be taken into account in order to assess the viability of this explanation. Therefore, we perform a model-independent global analysis in a Bayesian approach and find that the tension in the Cabibbo angle is significantly reduced, while the agreement with other data is also mostly improved. In fact, nonzero modifications of electron and muon neutrino couplings are preferred at more than 99.99\% C.L. (corresponding to more than $4\,\sigma$). Still, since constructive effects in the muon sector are necessary, simple models with right-handed neutrinos (whose global fit we update as a by-product) cannot fully explain data, pointing towards more sophisticated new physics models. 

\end{abstract}
\maketitle

\newpage
\section{Introduction}

The standard model (SM) of particle physics has been established with increasing precision within the last decades. In particular, both the electroweak (EW) fit~\cite{deBlas:2016ojx,Haller:2018nnx,Aaltonen:2018dxj} and the global fit~\cite{Ciuchini:2000de,Hocker:2001xe} of the Cabibbo-Kobayashi-Maskawa (CKM) matrix~\cite{Cabibbo:1963yz,Kobayashi:1973fv} are mainly in good agreement with the SM hypothesis and no new particles were directly observed at the LHC~\cite{Butler:2017afk,Masetti:2018btj}. Still, there are tensions between the different determinations of the Cabibbo angle from the CKM elements $V_{us}$ and $V_{ud}$ which became more pronounced recently. Here, $V_{us}$ from tau decays~\cite{Lusiani:2018ced,Amhis:2019ckw} and $V_{us}$ from kaon decays~\cite{Aoki:2019cca} do not perfectly agree. Furthermore, there is a $\sim 3-4\,\sigma$ tension between these determinations and the one from $V_{ud}$ entering super-allowed $\beta$ decay (using CKM unitarity) with non-negligible dependence on the theory predictions~\cite{Seng:2018yzq,Seng:2018qru,Gorchtein:2018fxl,Czarnecki:2019mwq}. In more detail, the different determinations of $V_{us}$ are as follows: (i) measurements of $K\to\pi\ell\nu$ together with the form factor $f_+(0)$ evaluated at zero momentum transfer result in $V_{us}=0.2232(11)$~\cite{Aoki:2019cca}. (ii) $K\to\ell\nu/\pi\to\ell\nu$ determines $V_{us}/V_{ud}$ once the ratio of decay constants $f_{K^{\pm}}/f_{\pi^{\pm}}$ is known. Using CKM unitarity this results in $V_{us}=0.22534(44)$~\cite{Aoki:2019cca}. (iii) $V_{ud}$ is measured via super-allowed nuclear $\beta$ decay. Here $V_{us}$ is again determined via CKM unitarity and using the theory input of Marciano {\it et al.}~\cite{Czarnecki:2019mwq} one finds $V_{us}=0.22699(77)$, while the evaluation of Seng {\it et al.} gives $V_{us}=0.22780(59)$~\cite{Seng:2018qru}. (iv) $V_{us}/V_{ud}$ is also measured in $\tau\to K\nu$, $\tau\to K\nu/\tau\to\pi\nu$ and via inclusive tau decays. Here the HFLAV average is $V_{us}=0.2221(13)$~\cite{Amhis:2019ckw}.

\begin{figure}[b]
	\includegraphics[width=0.45\textwidth]{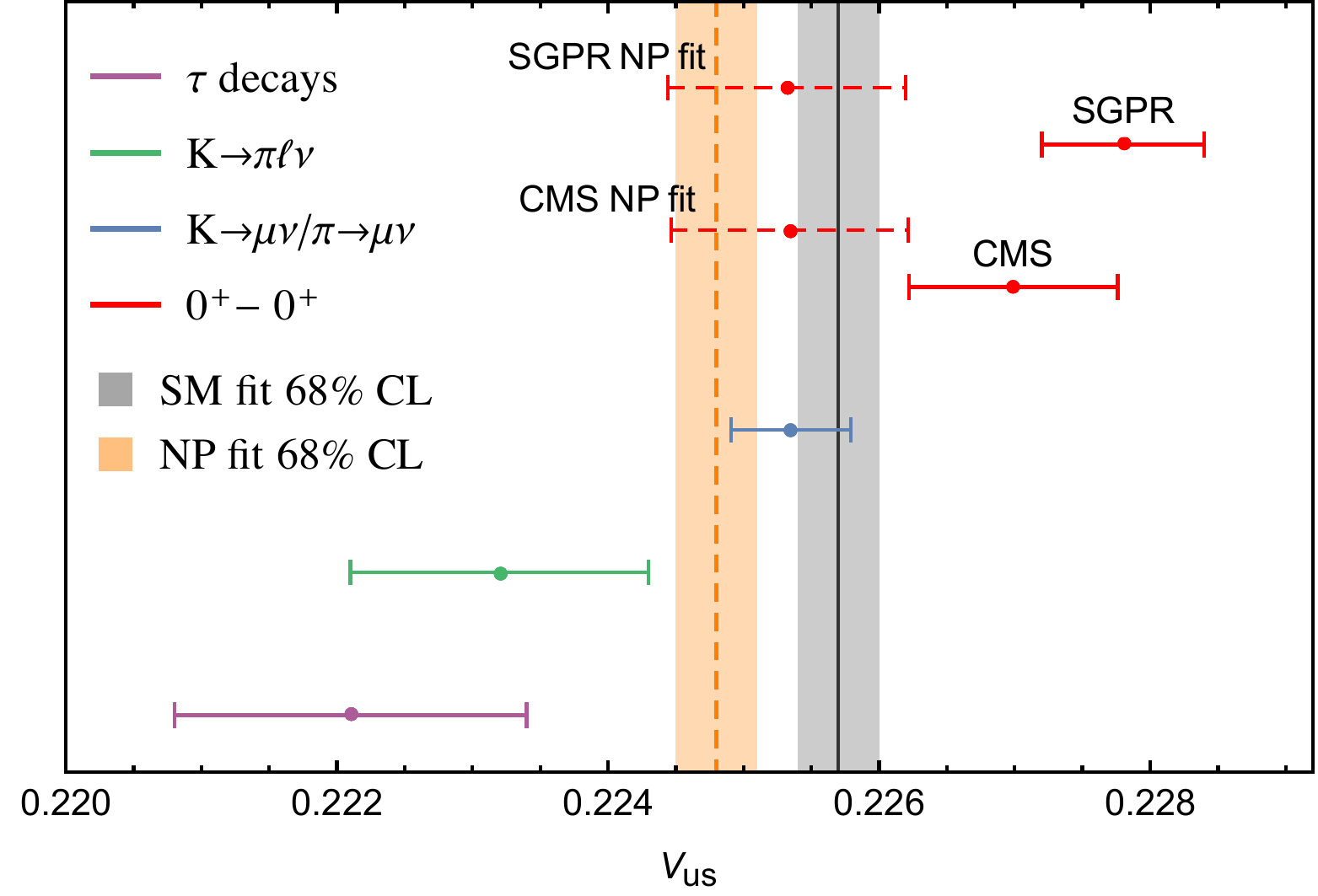}
	\caption{Measurements of $V_{us}$ from $\tau$ decays, $K\to \pi\ell\nu$, ${K\to\mu\nu}/{\pi\to\mu\nu}$, and $0^+ - 0^+$ transition using CKM unitarity to convert $V_{ud}$ to $V_{us}$. The grey band shows the 68\% C.L. posterior within the SM while the orange band corresponds to the NP fit with nonzero values of $\varepsilon_{ii}$. Here SGPR (CMS) stands for the $V_{us}$ values extracted from super-allowed beta decays using the theory input of Ref.~\cite{Seng:2018yzq} (Ref.~\cite{Czarnecki:2019mwq}). Accidentally, the posterior of $V_{us}$ is the same, independently of the theory input used for beta decays (to the numerical accuracy at which we are working). 
	\label{VusPlot}}
\end{figure}

This situation is graphically depicted in Fig.~\ref{VusPlot}. One can clearly see that these measurements are not consistent with each other, and Ref.~\cite{Grossman:2019bzp} quantifies this inconsistency to be at the level of $3.6\,\sigma$ ($5.1\,\sigma$) if the theory input of Ref.~\cite{Czarnecki:2019mwq} (Ref.~\cite{Seng:2018qru}) for super-allowed beta decay is used.

It is, therefore, very interesting to explore if new physics (NP) can explain this ``Cabibbo-angle anomaly.'' First of all, note that the absolute size of a NP effect potentially capable of explaining this anomaly is quite large since the corresponding SM contribution is generated at tree-level and is at most suppressed by one power of the Wolfenstein parameter. Because of this, at the level of effective operators, and given the strong LHC bounds on NP generating two-quark-two-lepton operators~\cite{Aaboud:2017buh}, NP entering via four-fermion operators seems to be a disfavoured option. Another possibility is a modification of $W$-fermion couplings, where a right-handed $W$-coupling to quarks only improves the fit mildly~\cite{Grossman:2019bzp}. Furthermore, a modification of left-handed $W$-couplings to quarks (which is equivalent to an apparent violation of CKM unitarity) can improve the agreement between super-allowed beta decay and $V_{us}$ from kaon decays~\cite{Belfatto:2019swo}, but generates potentially dangerous effects in other flavour observables (like kaon mixing). Therefore, we will follow a different and novel avenue in this Letter and study the impact of modified (flavour dependent) $W$-boson couplings to neutrinos.

Modified couplings of neutrinos to the SM $W$ are generated via higher dimensional operators in an EFT approach. Here, due to $SU(2)_L$ gauge invariance, in general not only $W$-neutrino couplings but also $Z$-neutrino couplings are modified. Moreover, these modified couplings not only enter $Z$ and $W$ decays, but also all low energy observables involving neutrinos. In particular, ratios testing lepton flavour universality (LFU) in $K$, $\pi$ and $W$ decays are most relevant due to their exquisite experimental and theoretical precision. There are stringent bounds from $K\to \mu\nu/K\to e\nu$~\cite{Ambrosino:2009aa,Lazzeroni:2012cx}, $\pi\to \mu\nu/\pi\to e\nu$~\cite{Aguilar-Arevalo:2015cdf,Tanabashi:2018oca} as well as from $\tau\to \mu\nu\nu/\tau\to e\nu\nu$ or $W\to\mu\nu/W\to e\nu$~\cite{Alcaraz:2006mx}. Correlated effects arise, and it is clear that a global fit to all data is necessary in order to assess consistently the impact of modified neutrino couplings. 

Modified neutrino couplings to SM gauge bosons have already been considered in the literature in the context of right-handed neutrinos~\cite{Lee:1977tib,Shrock:1980vy,Schechter:1980gr,Shrock:1980ct,Shrock:1981wq,Langacker:1988ur,Bilenky:1992wv,Nardi:1994iv,Tommasini:1995ii,Bergmann:1998rg,Loinaz:2002ep,Loinaz:2003gc,Loinaz:2004qc,Antusch:2006vwa,Antusch:2008tz,Biggio:2008in,Alonso:2012ji,Abada:2012mc,Akhmedov:2013hec,Basso:2013jka,Abada:2013aba,Antusch:2014woa,Antusch:2015mia,Abada:2015oba,Abada:2015trh,Abada:2016awd,Bolton:2019pcu} and global fits have also been performed~\cite{Fernandez-Martinez:2016lgt,Chrzaszcz:2019inj}. However, extensions with right-handed neutrinos lead necessarily to destructive interference with the SM, whereas here we will also be interested in the most general case allowing for an arbitrary phase of the NP contribution. The connection to, and correlations with, the Cabibbo-angle anomaly were not considered before and, in addition, we will use the publicly available \texttt{HEPfit} software~\cite{deBlas:2019okz} to perform a Bayesian analysis, while previous analyses were based on frequentist inference.

After defining our setup and reviewing the relevant observables together with the corresponding NP modification in the following section~\ref{Setup}, we will present the results of our fit in the analysis section before we conclude in the final section.

\section{Setup and observables}\label{Setup}

As outlined in the introduction, we want to assess the impact of modified neutrino couplings to gauge bosons within an EFT approach. For this purpose, we assume that the NP scale is above the EW scale, as suggested by LHC~\cite{Butler:2017afk,Masetti:2018btj} and LEP~\cite{Antonelli:2001rz} searches. Therefore, NP interactions must be $SU(2)_L$ gauge invariant and the number of operators is significantly reduced~\cite{Buchmuller:1985jz}. In fact, at the dimension 6 level, there is just one operator which modifies only the couplings of gauge bosons to neutrinos~\cite{Buchmuller:1985jz,Grzadkowski:2010es},
${\bar L_i}{\gamma^\mu }\tau _{}^I{L_j}{H^\dag }i{\mathord{\buildrel{\lower3pt\hbox{$\scriptscriptstyle\leftrightarrow$}} 
		\over D}_\mu^I}H$, with 	${\tau ^I} = \left( {1, - {\sigma _1}, - {\sigma _2}, - {\sigma _3}} \right)$,
where $\sigma_i$ are the Pauli matrices (i.e., it is the difference of the two operators $Q_{\varphi\ell}^{(1)}$ and $Q_{\varphi\ell}^{(3)}$ in the basis of Ref.~\cite{Grzadkowski:2010es}, to which we refer the interested reader for details on the conventions).
Note that this operator is Hermitian, meaning the diagonal elements are real. In what follows, we conveniently parametrize the effect of a nonzero Wilson coefficient of this operator in such a way that a neutrino entering a gauge coupling carries a (small) modification of $\frac{1}{2}\varepsilon_{ij}$, resulting in shifts in the $W$ and $Z$ Feynman rules,
\begin{align}
\frac{{ - i{g_2}}}{{\sqrt 2 }}{{\bar \ell }_i}{\gamma ^\mu }{P_L}{\nu _j}{W_\mu } &\Rightarrow \frac{{ - i{g_2}}}{{\sqrt 2 }}{{\bar \ell }_i}{\gamma ^\mu }{P_L}{\nu _j}{W_\mu }\left( {{\delta _{ij}} + \frac{1}{2}{\varepsilon _{ij}}} \right)\,,\nonumber\\
\frac{{ - i{g_2}}}{2c_W}{{\bar \nu }_i}{\gamma ^\mu }{P_L}{\nu _j}{Z_\mu } &\Rightarrow \frac{{ - i{g_2}}}{2c_W}{{\bar \nu }_i}{\gamma ^\mu }{P_L}{\nu _j}{Z_\mu }\left( {{\delta _{ij}} + {\varepsilon _{ij}}} \right)\,.
\label{couplings}
\end{align}
Here we assumed massless neutrinos and thus suppressed the PMNS matrix in the $W$ vertex.

Let us then, in the following subsections, consider the observables which will be included in our global fit.

\subsection{Lepton flavour violating decays}

Non-diagonal elements of $\varepsilon_{ij}$ lead to charged lepton flavour violation. Here the bounds from radiative lepton decays $\ell_i\to \ell_f\gamma$ are most stringent. Using the results of Ref.~\cite{Crivellin:2018qmi} we obtain
\begin{align}
\text{Br}\left(\ell_i\rightarrow \ell_f\gamma\right) = \frac{m_{\ell_i}^3}{4\pi\Gamma_{\ell_i}}\left|\frac{10e}{384\pi^2M_W^2}\frac{g_2^2}{2}m_{\ell_i}\varepsilon_{fi}\right|^2\,,
\end{align}
where we keep only linear terms in $\varepsilon_{if}$ and neglect the small mass of the outgoing lepton. The current experimental 90\% C.L. limits on lepton flavour violation processes are $4.2\times10^{-13}$ and $4.4(3.3)\times10^{-8}$ for $\mu\rightarrow e\gamma$~\cite{TheMEG:2016wtm} and $\tau\rightarrow \mu(e)\gamma$~\cite{Aubert:2009ag}, respectively, leading to $|\varepsilon_{e\mu}|\leq 1.4\times10^{-5}$ and $
|\varepsilon_{\mu\tau}|\leq 9.4\times10^{-3}$ and $
|\varepsilon_{e\tau}|\leq 1.1\times10^{-2}$. These limits on the flavour off-diagonal elements can be used directly, as they are unaffected (at leading order in $\varepsilon_{ij}$) by other entries $\varepsilon_{ij}$. Furthermore, since flavour off-diagonal elements of $\varepsilon_{ij}$ in flavour conserving processes do not interfere with the SM contributions, they enter only quadratically. Therefore, $\varepsilon_{ij}$ with $i\ne j$ can be safely neglected in the following observables.

\subsection{EW observables}

While the measurements of the mass of the $Z$ boson ($m_Z$) and the fine structure constant ($\alpha$) are not affected by the modification of the neutrino couplings in~\eq{couplings}, the Fermi constant ($G_F$, which is determined with a very high precision from the muon lifetime) is. As such, its value, extracted from $\mu\to e\nu\nu$, depends on the modification of the $W$-$\ell$-$\nu$ coupling. Taking into account that Br($\mu^+\rightarrow\e^+\nu_e\bar{\nu}_{\mu})\sim1$ we have
\begin{align}
\frac{1}{\tau_{\mu}}=\frac{(G_F^{\mathcal{L}})^2m_{\mu}^5}{192\pi^3}(1+\Delta q) \left(1+\frac{1}{2}\varepsilon_{ee}+\frac{1}{2}\varepsilon_{\mu\mu} \right)^2\,,
\end{align}
where $G_F^{\mathcal{L}}$ is the Fermi constant appearing in the Lagrangian and $\Delta q$ includes phase space, QED and hadronic radiative corrections. Thus we find
\begin{align}
\begin{split}
G_F^{}&=G_F^{\mathcal{L}} \left(1+\frac{1}{2}\varepsilon_{ee}+\frac{1}{2}\varepsilon_{\mu\mu}\right)\,.
\end{split}
\label{GFmod}
\end{align}
In addition to $G_F$, only the total width of the $Z$ ($\Gamma_Z$) and the number of light neutrino extracted from invisible $Z$ decays ($N_{\nu}$) receive direct modifications in the presence of anomalous neutrino couplings. The number of active neutrinos, as extracted from data~\cite{ALEPH:2005ab}, is given by
\begin{align}
\begin{split}
\text{N}_\nu^{\text{exp}}&=(1+\varepsilon_{ee})^2+(1+\varepsilon_{\mu\mu})^2+(1+\varepsilon_{\tau\tau})^2\\
&=2.9840\pm0.0082\,,
\end{split}
\end{align}
which in turn also changes $\Gamma_Z$, to which it contributes. 

We included the modifications of these observables into the EW implementations of \texttt{HEPfit}~\cite{deBlas:2019okz}; see the Supplemental Material~\cite{SM}, which includes 
Refs.~\cite{Schael:2013ita,Webber:2010zf,Aaboud:2018wps,
CMS:2019drq,TevatronElectroweakWorkingGroup:2016lid,
Aaboud:2018zbu,Sirunyan:2018mlv,Cirigliano:2007xi,Czapek:1993kc,
Britton:1992pg,Bryman:1982em,Antonelli:2010yf,Cirigliano:2011ny,
Awramik:2003rn,Sirlin:1980nh,Faisst:2003px,Avdeev:1994db,
Chetyrkin:1995ix,Chetyrkin:1995js,Broncano:2002rw,Abada:2007ux}, 
for further details.

\subsection{Test of LFU}

In case the diagonal elements of $\varepsilon_{ii}$ differ from each other, observables testing LFU provide stringent constraints. Here, we have ratios of $W$ decays ($W\to \ell_i\nu/W\to \ell_j\nu$) as well as of kaon, pion and tau decays (see Ref.~\cite{Pich:2013lsa} for an overview). Concerning $B$ decays, only $B\to D^{(*)}e\nu/B\to D^{(*)}\mu\nu$ provides a relevant constraint~\cite{Jung:2018lfu}. The corresponding observables, including their dependence on $\varepsilon_{ii}$ are shown in Table~\ref{LFUtest} of the Supplemental Material~\cite{SM}. For tau decays, we include their correlations as given in Ref.~\cite{Amhis:2019ckw}.

\subsection{Determination of $\mathbf{|V_{us}|}$}

We can now turn to the determination of $V_{us}$ as already briefly depicted in the introduction (see Fig.~\ref{VusPlot}).

$K_{\ell3}$: $V_{us}$ can be determined from the semi-leptonic kaon decays. In order to allow for LFU violation, one has to separate muon from electron modes. Averaging $K_L$, $K^\pm$ and $K_S$ modes~\cite{Tanabashi:2018oca}, one finds
\begin{align}
\begin{split}
 |V_{us}^{K_{\mu 3}}|&\simeq0.2234(8)\,,\qquad
 |V_{us}^{K_{e 3}}|\simeq0.2230(21)\,,
\end{split}
\end{align}
by using the lattice average~\cite{Aoki:2019cca} of the form factor at zero momentum transfer $f_+(0) = 0.9698(17)$,  $(N_f = 2+1+1)$. We choose to include the muon mode in the global fit, while the electron mode is already taken into account via the LFU ratios in Table~\ref{LFUtest} of the Supplemental Material~\cite{SM}. The NP modification, including the indirect effect of $G_F$, is
\begin{align}
|V_{us}^{K_{\mu 3}}|\simeq|V^{\mathcal{L}}_{us}|\left(1-\frac{1}{2}\varepsilon_{ee}\right)\,.
\end{align}

$K_{\ell2}$: $\text{Br}(K^{\pm}\rightarrow \mu^{\pm}\nu)/\text{Br}(\pi^{\pm}\rightarrow \mu^{\pm}\nu)$ determines $V_{us}/V_{ud}$. Including long-distance electromagnetic and strong isospin breaking corrections~\cite{Cirigliano:2011tm} and using the average of the lattice determinations for the ratio of form factors~\cite{Aoki:2019cca} $
{f_{K^\pm}}/{f_{\pi^{\pm}}} = 1.1967(18)$, with $(N_f = 2+1+1)$,
we find
 \begin{align}
 |V_{us}^{K/\pi}|\simeq0.22535(44)\, ,
 \end{align}
where we assumed CKM unitarity and took $|V_{ub}|\simeq0.004$. Note that the value of $V_{us}$ is very insensitive to $V_{ub}$, whose uncertainty can therefore be neglected, and that this $V_{ub}$ determination is not affected by $\varepsilon_{ij}$.

$0^+-0^+$ transitions: $|V_{ud}|$ can be extracted from super-allowed nuclear $\beta$ transitions~\cite{Hardy:2016vhg}. The result relies heavily on the evaluation of radiative corrections. We consider the two different results (as suggested in Ref.~\cite{Grossman:2019bzp}) of Marciano {\it et al.}~\cite{Czarnecki:2019mwq} (CMS) and Seng {\it et al.}~\cite{Seng:2018qru} (SGPR), which produce
\begin{align}
\begin{split}
&|V_{us}|_{\text{CMS}} =0.22699(77)\,, \qquad |V_{us}|_{\text{SGPR}} =0.22780(60)\,,\nonumber
\end{split}
\end{align}	
where CKM unitarity was again used. Turning on the NP couplings, we find the following modification
\begin{align}
|V_{us}^{\beta}|\simeq\sqrt{1-|V^{\mathcal{L}}_{ud}|^2\left(1-\frac{1}{2}\varepsilon_{\mu\mu}\right)^2}.
\end{align}

$\tau$ decays: $|V_{us}|$ can be also determined from hadronic $\tau$ decays~\cite{Amhis:2019ckw}. Here the average is~\cite{Amhis:2019ckw} 
$|V_{us}^\tau| = 0.2221(13)$.
Both $\tau\to K\nu/\tau\to\pi\nu$ and the inclusive mode measure $V_{us}/V_{ud}$, which means there is, at leading order, no dependence on $\varepsilon_{ij}$, and the determination is then unaffected by our NP contributions. This is different for the determination from $\tau\to K\nu$, whose dependence on $\varepsilon_{ij}$ is given by
\begin{equation}
|V_{us}^{\tau\to K\nu}| \simeq |V_{us}^{\mathcal{L}}|\left(1-\frac{1}{2}\varepsilon_{e e}-\frac{1}{2}\varepsilon_{\mu \mu}+\frac{1}{2}\varepsilon_{\tau \tau}\right).
\end{equation}
Since this mode as well as other hadronic tau decays are already included in the LFU ratios, we do not include the $V_{us}$ from tau decays in our global fit. Nevertheless, we can still predict the change in $V_{us}^{\tau\to K\nu}$.

\section{Analysis}
\label{analysis}

In this section we perform the global fit to the modified neutrino couplings [see \eq{couplings}], taking into account the observables discussed in the previous section. Before presenting the results, let us briefly discuss the statistical inference procedure we adopted. Our analysis is performed in a Bayesian framework using the publicly available \texttt{HEPfit} package~\cite{deBlas:2019okz}, whose Markov Chain Monte Carlo (MCMC) determination of posteriors is powered by the Bayesian Analysis Toolkit (\texttt{BAT})~\cite{Caldwell:2008fw}.

In order not to overweight the $V_{ud}$ measurements from $0^+-0^+$ transitions, we do not include both theory determinations at the same time, but rather define two scenarios: NP-I with $|V_{ud}|_{\text{CMS}}$ from Ref.~\cite{Czarnecki:2019mwq}, and
{NP-II} with $|V_{ud}|_{\text{SGPR}}$ from Ref.~\cite{Seng:2018yzq}.
Bayesian model comparison between different scenarios can be accomplished by evaluating an information criterion (IC)~\cite{Ando:2007,Ando:2011}.
This quantity is characterized by the mean and the variance of the posterior of the log-likelihood, $\log \mathcal{L}$, which yield an estimate of the predictive accuracy of the model~\cite{Gelman:2013}, and a penalty factor for the number of free parameters fitted. Preference for a model is given according to the smallest IC value, following the scale of evidence suggested in Refs.~\cite{Jeffreys:1998,Kass:1995}. The full list of fit parameters and details on the choice of priors can be found in Supplemental Material~\cite{SM}.

\begin{figure}[t]
	\includegraphics[width=0.45\textwidth]{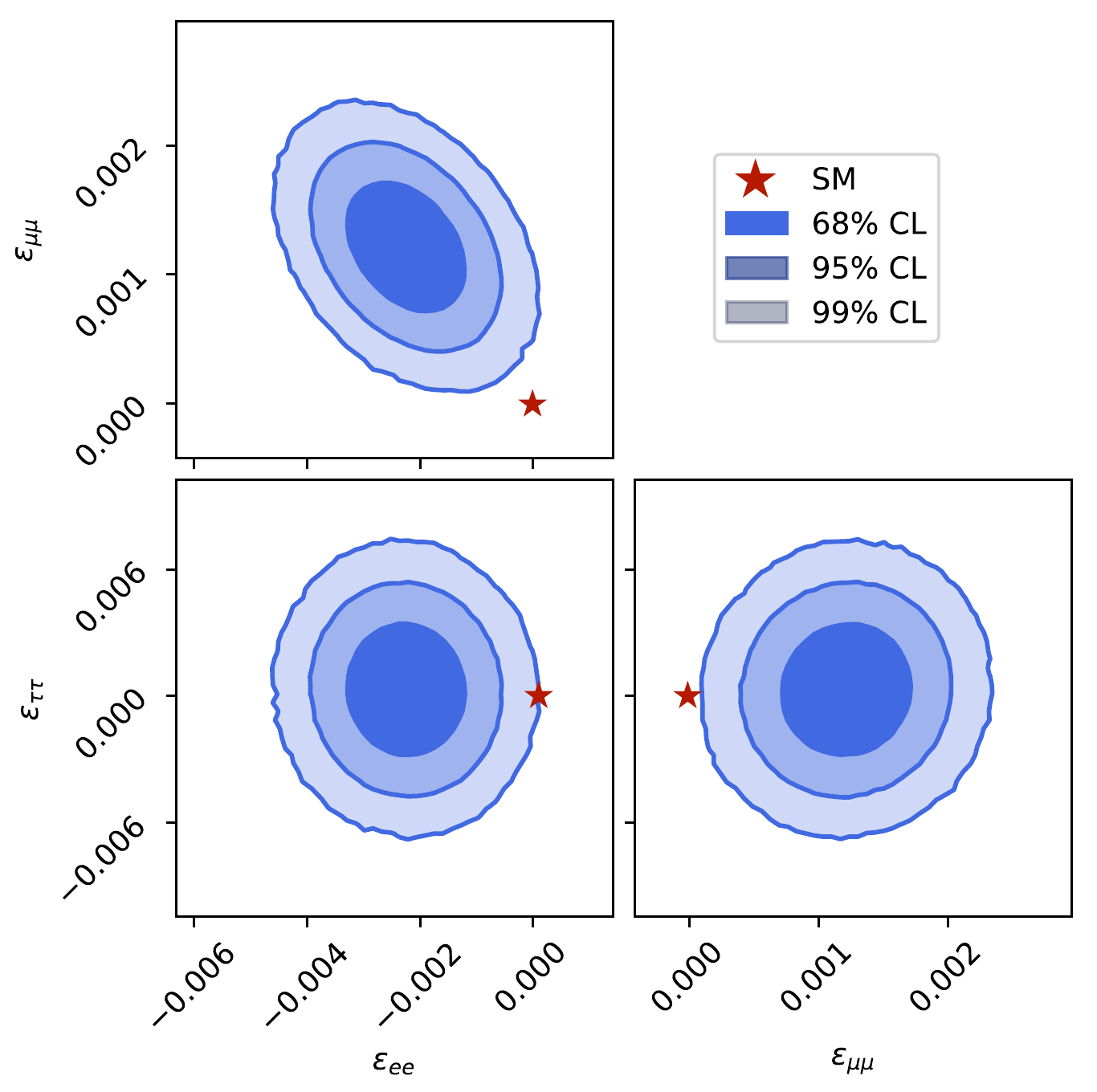}
	\caption{Global fit for scenario NP-II. The 2D fit ($68\%$, $95\%$ and $99\%$ C.L.) for $\varepsilon_{ii}$-$\varepsilon_{jj}$ as well as the 1D fit for each $\varepsilon_{ii}$ ($68\%$ C.L. indication) are shown.
	\label{Fig.fit}}
\end{figure}

Let us now probe the impact of nonzero values of $\varepsilon_{ij}$. As noted in the last section, one can neglect the flavour off-diagonal elements whose values are directly bounded by radiative lepton decays. As such, in the global fit we only have to consider $\varepsilon_{ee}$, $\varepsilon_{\mu\mu}$, and $\varepsilon_{\tau\tau}$. The 68\% C.L. intervals for fit parameters of the flavour sector ($V_{us}^{\mathcal L}$, $\varepsilon_{ii}$) within the two NP scenarios can be found in Table~\ref{Param} of the Supplemental Material~\cite{SM}. One can see that there is only a mild difference between both scenarios. In particular, the posterior of $V_{us}^{\mathcal L}$ is accidentally even the same and only the preferred region for $\varepsilon_{ee}$ ($\varepsilon_{\mu\mu}$) in scenario NP-II is slightly more negative (positive) than in scenario NP-I. Therefore, we only present the results for the two dimensional $\varepsilon_{ii}$-$\varepsilon_{jj}$ planes in Fig.~\ref{Fig.fit} within scenario NP-II (scenario NP-I is approximately $1\,\sigma$ more compatible with the SM hypothesis). There, the $68\%$, $95\%$ ad $99\%$ C.L. contours are shown, and it is clear from the $\varepsilon_{ee}$-$\varepsilon_{\mu\mu}$ plane, where the largest deviation from SM can be found, that these regions do not overlap with the SM point $\varepsilon_{ii}=0$, and that $\varepsilon_{ee}$ and $\varepsilon_{\mu\mu}$ possess an anti-correlation.

Concerning the $V_ {us}$ determination, we have also depicted the posterior of scenario NP-II and the updated values extracted from super-allowed beta decay in Fig.~\ref{VusPlot}. In summary, the main drivers leading to a better fit of the NP scenarios are the $V_{us}$ determination from super-allowed beta decays, $\tau\to\mu\nu\nu/\mu\to e\nu\nu$ and $\tau\to\mu\nu\nu/\tau\to e\nu\nu$~\cite{SM}.

For a more direct model comparison between the NP fits and the SM we look at the IC values. Here we obtain for the SM $\text{IC}_{\rm SM}=73$, compared to $\text{IC}_{\rm NP-II}=63$ and $\text{IC}_{\rm NP-I}=60$ for the two NP scenarios. In the vein of Ref.~\cite{Kass:1995}, this constitutes ``very strong'' evidence against the SM, further evidencing that current data clearly favours the NP hypothesis, and promoting the search for a NP model.

For that UV complete NP explanation obviously the possibility of right-handed neutrinos comes to mind, as these models give tree-level effects in $Z$-$\nu$-$\nu$ and $W$-$\ell$-$\nu$ couplings. However, here the effect is necessarily destructive (i.e., $\varepsilon_{ii}<0$), which is not in agreement with the preferred regions found in our fit. Nonetheless, performing the fit we find: $\varepsilon_{ee} = -0.0013^{+0.0006}_{-0.0006}$, $\varepsilon_{\tau \tau} = -0.0014^{+0.0010}_{-0.0015}$, and $0 > \varepsilon_{\mu\mu} > -0.0002$ at $68\%$ C.L. The shift towards values compatible with zero (within $\sim 2.2\,\sigma$ for $\varepsilon_{ee}$, and $1.2\,\sigma$ for $\varepsilon_{\tau\tau}$) signals a feeble improvement with respect to the SM. In fact, such a conclusion is supported by an IC value of 78, which is even bigger than the one of the SM due to the penalty for extra parameters. Moreover, once the constraint from $\mu\to e\gamma$, arising in models with $\varepsilon_{e\mu}=\sqrt{\varepsilon_{ee}\varepsilon_{\mu\mu}}$~\cite{Mohapatra:1986bd,Bernabeu:1987gr,Branco:1988ex,Buchmuller:1990du,Pilaftsis:1991ug,Dev:2012sg,Malinsky:2005bi,Antusch:2014woa,Coy:2018bxr} is taken into account, it is even more difficult in this scenario to explain data well.

Since right-handed neutrinos cannot fully explain the tensions within the EW fit, naturally the quest for a different UV completion arises. Even though a complete analysis is beyond the scope of this work, note that this can be achieved, e.g., by adding additional vectorlike leptons (VLLs) which induce tree-level modifications to $Z$ and $W$ couplings with leptons after EW symmetry breaking. Here, $SU(2)$ singlets and triplets generate the desired Wilson coefficients $C^{(1,3)}_{\phi \ell}$, while $SU(2)$ doublets modify couplings of right-handed charged leptons to $Z$. One has, therefore, four VLLs at our disposal: two doublets and two singlets which differ in hypercharge and contribute as
\begin{align}
\text{N}\!:\;&C_{\phi \ell}^{(3)fi}  = -C_{\phi \ell}^{(1)fi}<0,\;\;
\text{E}\!:\;C_{\phi \ell}^{(3)fi}  = C_{\phi \ell}^{(1)fi}<0,\nonumber\\
\Sigma_0\!:\;&C_{\phi \ell}^{(3)fi}  = \frac{1}{3}C_{\phi \ell}^{(1)fi}>0,\;\;
\Sigma_1\!:\; C_{\phi \ell}^{(3)fi}  = -\frac{1}{3}C_{\phi \ell}^{(1)fi}>0,\nonumber
\end{align}   
using the conventions of Ref.~\cite{deBlas:2017xtg}. It is thus clear that we can produce any combination of $C^{(1,3)}_{\phi \ell}$ with arbitrary sign, including $C^{(1)}_{\phi \ell}=- C^{(3)}_{\phi \ell}$ being positive or negative. Such a linear combination of vectorlike leptons might seem {\it ad hoc} at first glance. However, this is exactly what happens in composite or extradimensional models with custodial protection~\cite{Agashe:2003zs,Agashe:2006at} [see e.g. Refs.~\cite{delAguila:2010vg,Carmona:2013cq} for a generalization to the lepton sector with $SU(2)_L$ triplets]. In such models, the symmetry group is chosen in such a way that the VLL representations (generated for instance as Kaluza-Klein excitations) lead to modifications of $Z$-$\nu$-$\nu$ and  $W$-$\ell$-$\nu$ couplings, but not to $Z$-$\ell$-$\ell$ interactions. As this corresponds exactly to the case at hand, $C^{(1)}_{\phi \ell}=- C^{(3)}_{\phi \ell}$, extradimensional or composite models with custodial protection can therefore very well give rise to the scenario obtained in our model-independent fit.

\section{Conclusions and Outlook}\label{Conclusions}

In this Letter, we performed a model-independent global fit to modified neutrino couplings motivated by the Cabbibo-angle anomaly (i.e., the disagreement between the different determinations of $V_{us}$). Taking into account all relevant observables related to the EW sector of the SM and observables testing LFU (like $\tau\to\mu\nu\nu/\tau\to e\nu\nu$, $\pi\to\mu\nu\nu/\pi\to e\nu\nu$, etc.), we found that agreement with data can be significantly improved by small modifications $\varepsilon_{ii}$. Our results for this NP scenario are depicted in Fig.~\ref{Fig.fit}, showing the SM hypothesis lies beyond the 99.99\% C.L. region, corresponding to a deviation of more than $4\,\sigma$. Furthermore, the IC values of the scenarios here considered strongly prefer the NP hypothesis. 

However, conventional models with right-handed neutrinos, which lead to necessary destructive interference, cannot explain data very well. Nevertheless, since these models are well motivated by the observed nonvanishing neutrino masses, we updated their global fit, taking into account the different $V_{us}$ determinations.

Clearly, more data and further theory input is needed to clarify the situation in the future. Also, the study and construction of NP models which can give a constructive effect in $Z$-$\nu$-$\nu$ and $W$-$\ell$-$\nu$ couplings, in particular strongly coupled theories with custodial protection, is a promising direction of research, building upon the results of this article. Furthermore, as our explanation involves flavour-dependent couplings, the Cabibbo-angle anomaly fits into the bigger picture of deviations from LFU as observed in $b\to s\ell^+\ell^-$ transitions~\cite{Capdevila:2017bsm,Altmannshofer:2017yso,DAmico:2017mtc,Ciuchini:2017mik,Hiller:2017bzc,Geng:2017svp,Hurth:2017hxg,Alguero:2019ptt,Aebischer:2019mlg,Ciuchini:2019usw,Arbey:2019duh} and the anomalous magnetic moment of the muon and electron~\cite{Davoudiasl:2018fbb,Crivellin:2018qmi}. This opens up the possibility of so far undiscovered correlations among these observables with UV complete models.
\medskip

{\it Acknowledgements} --- {\small
	We thank Marco Fedele, Julian Heeck, Martin Hoferichter, Ayan Paul, Hugo M. Proença and Mauro Valli for useful discussions and/or help with \texttt{HEPfit}. The work of A.C. is supported by a Professorship Grant (PP00P2\_176884) of the Swiss National Science Foundation. A.M.C. acknowledges support by the Swiss National Science Foundation under contract 200021\_178967.}
	
\appendix


\section*{Supplemental Material}

\subsection{Electroweak observables}

\begin{table}
	\resizebox{0.5\textwidth}{!}{		\begin{tabular}{l c c }
			\hline\hline
			Observable & Ref. & Measurement \\
			\hline
			$\frac{K\rightarrow\mu\nu}{K\rightarrow e\nu}\simeq|1+\frac{1}{2}\varepsilon_{\mu\mu}-\frac{1}{2}\varepsilon_{ee}|$ &~\cite{Pich:2013lsa} &$0.9978 \pm 0.0020$ \\		
			$\frac{\pi\rightarrow\mu\nu}{\pi\rightarrow e\nu}\simeq|1+\frac{1}{2}\varepsilon_{\mu\mu}-\frac{1}{2}\varepsilon_{ee}|$&~\cite{Aguilar-Arevalo:2015cdf,Tanabashi:2018oca} & $1.0010 \pm 0.0009$ \\		
			$\frac{\tau\rightarrow\mu\nu\bar{\nu}}{\tau\rightarrow e\nu\bar{\nu}}\simeq|1+\frac{1}{2}\varepsilon_{\mu\mu}-\frac{1}{2}\varepsilon_{ee}|$&~\cite{Amhis:2019ckw,Tanabashi:2018oca} & $1.0018 \pm 0.0014$ \\		
			$\frac{K\rightarrow\pi\mu\bar{\nu}}{K\rightarrow\pi e\bar{\nu}}\simeq|1+\frac{1}{2}\varepsilon_{\mu\mu}-\frac{1}{2}\varepsilon_{ee}|$&~\cite{Pich:2013lsa} & $1.0010 \pm 0.0025$ \\		
			$\frac{W\rightarrow\mu\bar{\nu}}{W\rightarrow e\bar{\nu}}\simeq|1+\frac{1}{2}\varepsilon_{\mu\mu}-\frac{1}{2}\varepsilon_{ee}|$&~\cite{Pich:2013lsa,Schael:2013ita} & $0.996 \pm 0.010$ \\	
			$\frac{\tau\rightarrow e\nu\bar{\nu}}{\mu\rightarrow e\bar{\nu}\nu}\simeq|1+\frac{1}{2}\varepsilon_{\tau\tau}-\frac{1}{2}\varepsilon_{\mu\mu}|$&~\cite{Amhis:2019ckw,Tanabashi:2018oca} & $1.0010 \pm 0.0014$ \\			
			$\frac{\tau\rightarrow \pi\nu}{\pi\rightarrow \mu\bar{\nu}}\simeq|1+\frac{1}{2}\varepsilon_{\tau\tau}-\frac{1}{2}\varepsilon_{\mu\mu}|$&~\cite{Amhis:2019ckw}& $0.9961 \pm 0.0027$ \\			
			$\frac{\tau\rightarrow K\nu}{K\rightarrow \mu\bar{\nu}}\simeq|1+\frac{1}{2}\varepsilon_{\tau\tau}-\frac{1}{2}\varepsilon_{\mu\mu}|$&~\cite{Amhis:2019ckw} & $0.9860 \pm 0.0070$ \\			
			$\frac{W\rightarrow \tau\bar{\nu}}{W\rightarrow \mu\bar{\nu}}\simeq|1+\frac{1}{2}\varepsilon_{\tau\tau}-\frac{1}{2}\varepsilon_{\mu\mu}|$&~\cite{Pich:2013lsa,Schael:2013ita} & $1.034 \pm 0.013$ \\			
			$\frac{\tau\rightarrow \mu\nu\bar{\nu}}{\mu\rightarrow e\nu\bar{\nu}}\simeq|1+\frac{1}{2}\varepsilon_{\tau\tau}-\frac{1}{2}\varepsilon_{ee}|$&~\cite{Amhis:2019ckw,Tanabashi:2018oca} & $1.0029 \pm 0.0014$ \\			
			$\frac{W\rightarrow \tau\bar{\nu}}{W\rightarrow e\bar{\nu}}\simeq|1+\frac{1}{2}\varepsilon_{\tau\tau}-\frac{1}{2}\varepsilon_{ee}|$&~\cite{Pich:2013lsa,Schael:2013ita} & $1.031 \pm 0.013$\\
			$\frac{B\rightarrow D^{(*)}\mu\nu}{B\rightarrow D^{(*)}e\nu}\simeq|1+\frac{1}{2}\varepsilon_{\mu\mu}-\frac{1}{2}\varepsilon_{ee}|$&~\cite{Jung:2018lfu} & $0.989 \pm 0.012$\\
			\hline\hline
	\end{tabular}}	\caption{Ratios testing LFU together with their dependence on the modifications of the neutrino couplings $\varepsilon_{ij}$. Here it is implicitly understood that these ratios are normalized to their values within the SM, such that any deviation from unity measures LFU violation.}\label{LFUtest}
\end{table}

Measurements of the EW observables, as performed at LEP~\cite{Schael:2013ita,ALEPH:2005ab}, are high precision tests of the SM. The EW sector of the SM can be completely parametrized by the three Lagrangian parameters $v$, $g_1$ and $g_2$; then, other quantities like $G_F$, $m_W$ or $m_Z$ can be expressed in terms of these parameters and their measurements allow for consistency tests. However, for practical purposes it is better to choose another set of three parameters parametrizing the EW sector of the SM: a convenient choice is to use the quantities with the smallest experimental error of their direct measurements, i.e. the mass of the $Z$ boson ($m_Z$), the Fermi constant ($G_F$) and the fine structure constant ($\alpha$).

The EW observables, computed from $G_F$, $m_Z$ and $\alpha$, are given at the beginning of Table~\ref{Obs}. Since the $Z$ sector remains lepton flavour universal (for charged leptons) we can thus use the standard $Z$-pole observables (assuming LFU)~\cite{ALEPH:2005ab}. The Higgs mass ($m_h$), the top mass ($m_t$) and the strong coupling constant ($\alpha_s$) have to be included as fit parameters as well, since they enter indirectly EW observables via loop effects.

We point out that in our analysis it is convenient that we use $G_F$, not $G_F^{\mathcal{L}}$, as a fit parameter whose prior is determined by its direct measurement $1.1663787(6)\times10^{-5}\,\text{GeV}^{-2}$~\cite{Webber:2010zf}. Note that $G_F^{\mathcal{L}}$ enters all other EW observables which are thus indirectly modified by $\varepsilon_{ee}$ and $\varepsilon_{\mu\mu}$. 

\begin{table}[t!]
	\centering
	\resizebox{0.49\textwidth}{!}{	\begin{tabular}{l c l c c c c}
			\hline\hline
			Parameter && Prior && SM posterior \\
			\hline
			$G_F\,\,[{\rm GeV}^{-2}]$~\cite{Tanabashi:2018oca} && $1.1663787(6) \times 10^{-5}$ &&$\star$ \\
			$\alpha$~\cite{Tanabashi:2018oca} && $7.2973525664(17) \times 10^{-3}$  && $\star$\\
			$\Delta\alpha_{\rm had}$~\cite{Tanabashi:2018oca} && $276.1(11) \times 10^{-4}$  && $275.4(10) \times 10^{-4}$\\
			$\alpha_s(M_Z)$~\cite{Tanabashi:2018oca} && $0.1181(11)$ && $\star$\\
			$m_Z\,\,[{\rm GeV}]$~\cite{ALEPH:2005ab} && $91.1875\pm0.0021$ &&$91.1883\pm0.0020$ \\
			$m_H\,\,[{\rm GeV}]$~\cite{Aaboud:2018wps,CMS:2019drq} && $125.16\pm0.13$ &&$\star$ \\
			$m_{t}\,\,[{\rm GeV}]$~\cite{TevatronElectroweakWorkingGroup:2016lid,Aaboud:2018zbu,Sirunyan:2018mlv} && $172.80\pm0.40$  &&$172.96\pm0.39$ \\
			\hline\hline
	\end{tabular}}
	\caption{Parameters of the EW fit together with their (Gaussian) priors and posteriors. Here, the posteriors that remain equal to the prior (up to the numerical accuracy considered) are abbreviated by a star. Furthermore, once the effect of $\varepsilon_{ij}$ is included, only the posterior of the top (pole) mass changes slightly to $m_t\sim 172.83\,$GeV while all other parameters remain the same, showing the EW fit is, to a very good approximation, independent of the flavour parameters. \label{ParamEW}}
\end{table}

\begin{table}[t!]
	\centering
	\begin{tabular}{l | c c c c c c c c c c c}
		&&  Prior &&
		NP-I posterior && 
		NP-II posterior	\\	\hline
		$V_{us}^{\mathcal{L}}$&& $0.225\pm0.010$ &&$0.2248\pm0.0004$ && $0.2248\pm0.0004$\\
		$\varepsilon_{ee}$&& $0.00\pm0.05$  &&$-0.0018\pm0.0006 $&&$-0.0022\pm0.0007$ \\
		$\varepsilon_{\mu\mu}$ &&  $0.00\pm0.05$ &&$0.0008\pm0.0004$ && $0.0012\pm0.0003$\\
		$\varepsilon_{\tau\tau}$&& $0.00\pm0.05$ &&$-0.0002\pm0.0020$ && $-0.0003\pm0.0020$\\
	\end{tabular}
	\caption{Fit parameters of the flavour sector together with their posteriors for the two NP scenarios. One can see that the choice between the two $V_{us}$ determinations has a small effect on $\varepsilon_{ee}$ and $\varepsilon_{\tau\tau}$, while in scenario NP-I a vanishing value of $\varepsilon_{\mu\mu}$ is more compatible with data (within $\sim 2~\sigma$). Note that the value for $V_{us}^{\mathcal L}$ is accidentally the same (to the numerical accuracy given): the effect of the different $V_{us}$ determinations is compensated by the difference in the preferred regions for $\varepsilon_{ii}$. However, the deviation from the SM posterior, $0.2257(3)$, is indeed sizable. \label{Param}}
\end{table}

\begin{table*}[t!]
	\begin{tabular}{l c l c c c c c c c c c c c c } \toprule
		Observable & Ref. & Measurement & SM Posterior & NP-I posterior & NP-II posterior& Pull I & Pull II \\
		\colrule
		$M_W\,[\text{GeV}]$ & ~\cite{Tanabashi:2018oca} & $80.379(12)$ & $80.363(4)$ & $80.371(6)$ & $80.370(6)$ & \cellcolor{Gray} 0.67  & \cellcolor{Gray} 0.59  \\
		$\Gamma_W\,[\text{GeV}]$ & ~\cite{Tanabashi:2018oca} & $2.085(42)$ & $2.089(1)$ & $2.090(1)$ & $2.090(1)$ &  -0.02 &   -0.02\\
		$\text{BR}(W\to \text{had})$ & ~\cite{Tanabashi:2018oca} & $0.6741(27)$ & $0.6749(1)$ & $0.6749(2)$  & $0.6749(1)$ & 0 & 0 \\
		$\text{sin}^2\theta_{\rm eff}^{\rm lept}(Q^{\rm had}_{\rm FB})$& ~\cite{Tanabashi:2018oca}  & $0.2324(12)$ &$0.2316(4)$ & $0.2315(1)$ & $0.2315(1)$  & -0.1 & -0.1\\
		$\text{sin}^2\theta_{\rm eff(Tev)}^{\rm lept}$ & ~\cite{Tanabashi:2018oca}  & $0.23148(33)$ &$0.2316(4)$  & $0.2315(1)$  & $0.2315(1)$  & \cellcolor{Gray} 0.17 & \cellcolor{Gray} 0.17\\
		$\text{sin}^2\theta_{\rm eff(LHC)}^{\rm lept}$ & ~\cite{Tanabashi:2018oca}  & $0.23104(49)$ &$0.2316(4)$ & $0.2315(1)$  & $0.2315(1)$  & -0.03 &-0.03 \\
		$P_{\tau}^{\rm pol}$ &~\cite{ALEPH:2005ab} &$0.1465(33)$ & $0.1461(3)$ & $0.1474(8)$  & $0.1472(8)$& -0.14 & -0.09\\
		$A_{\ell}$ &~\cite{ALEPH:2005ab} &$0.1513(21)$ & $0.1461(3)$ & $0.1474(8)$ & $0.1472(8)$ & \cellcolor{Gray} 0.72 & \cellcolor{Gray} 0.60 \\
		$\Gamma_Z\,[\text{GeV}]$ &~\cite{ALEPH:2005ab} &$2.4952(23)$ & $2.4947(6)$  &$2.496(1)$ & $2.496(1)$ & -0.11  & -0.11  \\
		$\sigma_h^{0}\,[\text{nb}]$ &~\cite{ALEPH:2005ab} &$41.541(37)$ & $41.485(6)$ & $41.495(24)$ & $41.493(24)$& \cellcolor{Gray} 0.47 &\cellcolor{Gray} 0.42 \\
		$R^0_{\ell}$ &~\cite{ALEPH:2005ab} &$20.767(35)$ & $20.747(7)$ & $20.749(7)$ & $20.749(7)$ & \cellcolor{Gray} 0.06 & \cellcolor{Gray} 0.06\\
		$A_{\rm FB}^{0,\ell}$&~\cite{ALEPH:2005ab} &$0.0171(10)$ & $0.0160(7)$ & $0.0163(2) $ & $0.0163(2) $ & \cellcolor{Gray} 0.12 & \cellcolor{Gray} 0.12  \\
		$R_{b}^{0}$ &~\cite{ALEPH:2005ab} &$0.21629(66)$ & $0.21582(1)$ & $0.21582(1)$ & $0.21582(1)$ &  0 & 0\\
		$R_{c}^{0}$ &~\cite{ALEPH:2005ab} &$0.1721(30)$ &$0.17219(2)$ & $0.17220(2)$ & $0.17220(2)$ & 0 & 0\\
		$A_{\rm FB}^{0,b}$ &~\cite{ALEPH:2005ab} &$0.0992(16)$ & $0.1024(2)$ & $0.1033(6)$ & $0.1032(6)$ &  -0.41& -0.36 \\
		$A_{\rm FB}^{0,c}$ &~\cite{ALEPH:2005ab} &$0.0707(35)$ & $0.0731(2)$ & $0.0738(4)$ & $0.0738(4)$ & -0.20 &  -0.20\\
		$A_{b}$ &~\cite{ALEPH:2005ab} &$0.923(20)$ & $0.93456(2)$ & $0.9347(1)$ & $0.9347(1)$ & -0.01 & -0.01\\
		$A_{c}$ &~\cite{ALEPH:2005ab} &$0.670(27)$ & $0.6675(1)$  & $0.6680(4)$ & $0.6680(3)$ & 0 & 0\\
		\colrule \\[-0.30cm]
		$\frac{K\rightarrow\mu\nu}{K\rightarrow e\nu}$ &~\cite{Lazzeroni:2012cx,Ambrosino:2009aa,Cirigliano:2007xi,Pich:2013lsa} &$0.9978 \pm 0.0020$ & $1$ &$1.00137\pm0.00046$ &$1.00173\pm0.00043$ &-0.63 &-0.82\\
		$\frac{\pi\rightarrow\mu\nu}{\pi\rightarrow e\nu}$&~\cite{Czapek:1993kc,Britton:1992pg,Bryman:1982em, Cirigliano:2007xi,Aguilar-Arevalo:2015cdf,Tanabashi:2018oca} & $1.0010 \pm 0.0009$ &$1$ & $1.00137\pm0.00046$&$1.00173\pm0.00043$ &\cellcolor{Gray} 0.75&\cellcolor{Gray} 0.38\\
		$\frac{\tau\rightarrow\mu\nu\bar{\nu}}{\tau\rightarrow e\nu\bar{\nu}}$&~\cite{Amhis:2019ckw,Tanabashi:2018oca} & $1.0018 \pm 0.0014$ & $1$& $1.00137\pm0.00046$&$1.00173\pm0.00043$ &\cellcolor{Gray} 0.99&\cellcolor{Gray} 1.24\\
		$\frac{K\rightarrow\pi\mu\bar{\nu}}{K\rightarrow\pi e\bar{\nu}}$&~\cite{Antonelli:2010yf,Cirigliano:2011ny,Pich:2013lsa} & $1.0010 \pm 0.0025$ &$1$ & $1.00137\pm0.00046$&$1.00173\pm0.00043$ &\cellcolor{Gray} 0.25& \cellcolor{Gray} 0.11\\
		$\frac{W\rightarrow\mu\bar{\nu}}{W\rightarrow e\bar{\nu}}$&~\cite{Pich:2013lsa,Schael:2013ita} & $0.996 \pm 0.010$ &$1$ &$1.00137\pm0.00046$ &$1.00173\pm0.00043$ &-0.14& -0.17\\
		$\frac{B\rightarrow D^{(*)}\mu\nu}{B\rightarrow D^{(*)}e\nu}$ &~\cite{Jung:2018lfu} & $0.989\pm0.012$ & $1$ & $1.00137\pm0.00046$ &$1.00173\pm0.00043$ &-0.11&-0.14 \\
		$\frac{\tau\rightarrow e\nu\bar{\nu}}{\mu\rightarrow e\bar{\nu}\nu}$&~\cite{Amhis:2019ckw,Tanabashi:2018oca} & $1.0010 \pm 0.0014$ &$1$& $0.9997\pm0.0010$&$0.9995\pm0.0010$ &-0.04&-0.15\\
		$\frac{\tau\rightarrow \pi\nu}{\pi\rightarrow \mu\bar{\nu}}$&~\cite{Amhis:2019ckw}& $0.9961 \pm 0.0027$ &$1$&$0.9997\pm0.0010$ &$0.9995\pm0.0010$ &\cellcolor{Gray} 0.20& \cellcolor{Gray} 0.26\\
		$\frac{\tau\rightarrow K\nu}{K\rightarrow \mu\bar{\nu}}$&~\cite{Amhis:2019ckw} & $0.9860 \pm 0.0070$ &$1$& $0.9997\pm0.0010$&$0.9995\pm0.0010$ &\cellcolor{Gray} 0.06&\cellcolor{Gray} 0.09\\
		$\frac{W\rightarrow \tau\bar{\nu}}{W\rightarrow \mu\bar{\nu}}$&~\cite{Pich:2013lsa,Schael:2013ita} & $1.034 \pm 0.013$ &$1$&$0.9997\pm0.0010$ &$0.9995\pm0.0010$ &-0.02&-0.03\\
		$\frac{\tau\rightarrow \mu\nu\bar{\nu}}{\mu\rightarrow e\nu\bar{\nu}}$&~\cite{Amhis:2019ckw,Tanabashi:2018oca} & $1.0029 \pm 0.0014$ &$1$& $1.0011\pm0.0011$&$1.0013\pm0.0011$ &\cellcolor{Gray}1.06&\cellcolor{Gray}1.17\\
		$\frac{W\rightarrow \tau\bar{\nu}}{W\rightarrow e\bar{\nu}}$&~\cite{Pich:2013lsa,Schael:2013ita} & $1.031 \pm 0.013$ &$1$&$1.0011\pm0.0011$ &$1.0013\pm0.0011$ &\cellcolor{Gray}0.10& \cellcolor{Gray}0.11\\[0.05cm]
		\colrule \\[-0.30cm]
		$|V_{us}^{K_{\mu 3}}|$&~\cite{Tanabashi:2018oca,Aoki:2019cca} & $0.2234 \pm 0.0008$ &$0.2257(3)$&$0.22509 \pm 0.00040$ &$0.22516 \pm 0.00040$ &\cellcolor{Gray}0.81&\cellcolor{Gray} 0.74\\
		$|V_{us}/V_{ud}|^{K/\pi}$&~\cite{Cirigliano:2011tm,Aoki:2019cca} & $0.2313 \pm 0.0005$ & $0.2317(4)$ & $0.23078 \pm 0.00044$ &$0.23082 \pm 0.00044$ &-0.16&-0.10\\	
		$|V_{us}^{\tau}|_{\text{incl.}}$ &~\cite{Hardy:2016vhg,Czarnecki:2019mwq} & $0.2195 \pm 0.0019$ &  $0.2257(3)$ &$0.22487\pm0.00041$ & $0.22491\pm0.00041$ &\cellcolor{Gray} 0.48 &\cellcolor{Gray} 0.45\\
		$|V_{ud}^{\beta}|_{\text{CMS}}$ &~\cite{Hardy:2016vhg,Czarnecki:2019mwq} & $0.97389 \pm 0.00018$ & $0.974185(79)$  &$0.97400\pm 0.00017$  & - &\cellcolor{Gray}0.56&-\\
		$|V_{ud}^{\beta}|_{\text{SGPR}}$&~\cite{Hardy:2016vhg,Seng:2018qru} & $0.97370 \pm 0.00014$ & $0.974185(79)$ &-&$0.97379\pm0.00013$  &-&\cellcolor{Gray}2.57\\
		\botrule
	\end{tabular}
	\caption{Observables included in and predicted by our global fit. Here the SM scenario, as well as the NP-I and NP-II hypotheses are shown. The pulls are defined with respect to the SM in such a way that positive values means better agreement of the NP hypothesis (highlighted in grey). Note that the observables testing LFU are implicitly normalized in such a way that they correspond to unity within the SM.\label{Obs}}
\end{table*}

\subsection{Fit results}

Employing the Metropolis-Hastings algorithm implemented in \texttt{BAT} to sample from the posterior distribution, our MCMC runs involved 6 chains with a total of 2 million events per chain, collected after an equivalent number of pre-run iterations.

The fit parameters of the EW sector, $G_F,\,\alpha,\,\alpha_s,\,M_Z,\,m_t$ and $m_H$, are reported in Table~\ref{ParamEW}. Here we assume Gaussian priors, which correspond to the current direct measurements or evaluations of these parameters. We verified that the chosen ranges of flat priors yield well-determined probability density functions (\textit{p.d.f.}), i.e. they are chosen in such a way that larger ranges would not significantly alter our results.

Concerning the $W$ mass computation, \texttt{HEPfit} provides both the option of using the recent precise numerical formula from Ref.~\cite{Awramik:2003rn}, and the usual determination of $M_W$ from the $Z$-boson mass, $\alpha$, and $G_F$~\cite{Sirlin:1980nh}, with radiative corrections encoded in $\Delta r$ (which has been known up to 3 loop $\mathcal{O}(\alpha^3)$ EW~\cite{Faisst:2003px} and $\mathcal{O}(\alpha \alpha_s^2,\, \alpha^2 \alpha_s)$ EW-QCD contributions~\cite{Faisst:2003px,Avdeev:1994db,Chetyrkin:1995ix,Chetyrkin:1995js}). Due to the direct modifications of $G_F$ under analysis, we opted for the latter.
 
 \begin{figure*}[t!]
 	\includegraphics[width=0.4\textwidth]{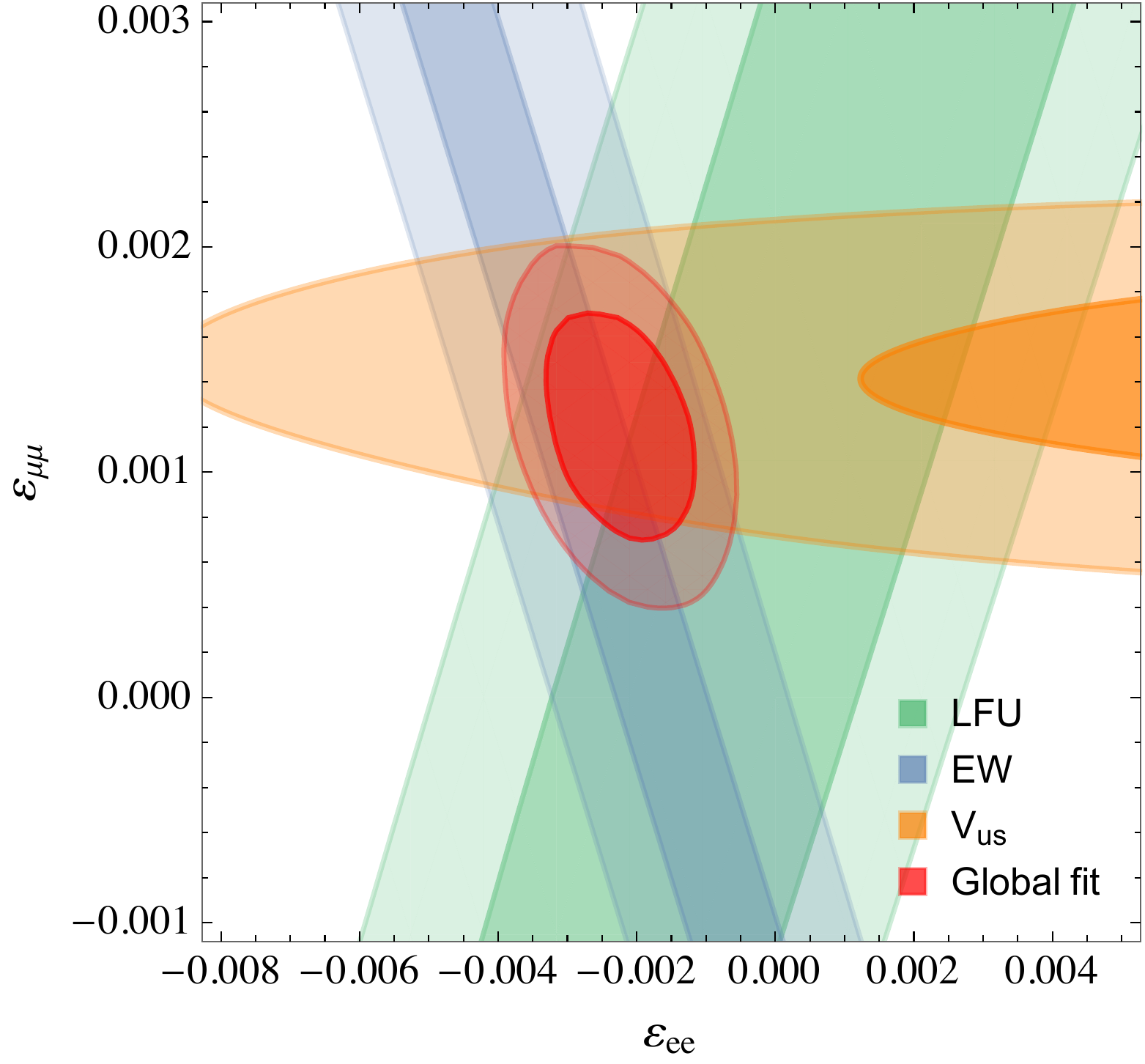}\hspace{1 cm}
 		\includegraphics[width=0.5\linewidth]{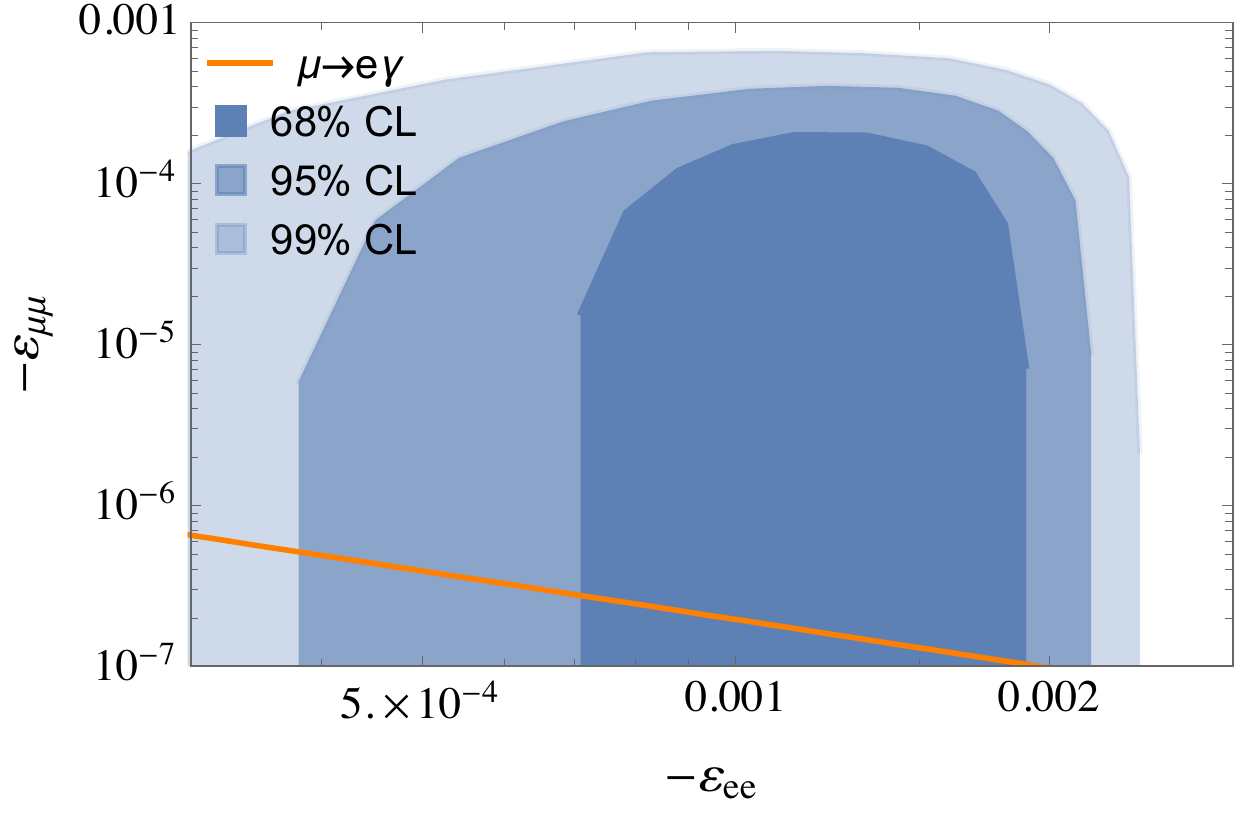}
 	\caption{Left: Contributions to the global fit from different classes of observables: LFU (green), EW (blue) and $V_{us}$ (orange). The global fit (red) is superimposed. Dark and light colors correspond to the 68\%~CL and 95\%~CL, respectively.
 		Right: Global fit for scenario NP-II confining $\varepsilon_{ii}$ to be negative, as obtained in models with effective PMNS unitarity violation from mixing with right-handed neutrinos~\cite{Broncano:2002rw,Abada:2007ux,Antusch:2006vwa,Antusch:2014woa}and considered in previous fits~\cite{Fernandez-Martinez:2016lgt,Chrzaszcz:2019inj}. In addition, we show the 90\%CL region from $\mu\to e\gamma$ as obtained in simple models with approximate symmetry protecting neutrino masses $\varepsilon_{e\mu}=\sqrt{\varepsilon_{ee}\varepsilon_{\mu\mu}}$~\cite{Mohapatra:1986bd,Bernabeu:1987gr,Branco:1988ex,Buchmuller:1990du,Pilaftsis:1991ug,Dev:2012sg,Malinsky:2005bi,Antusch:2014woa,Coy:2018bxr}\label{FitReg}}
 \end{figure*}
 
First, we redo the global EW fit within the SM; then, we probe the impact of non-zero values of $\varepsilon_{ij}$. The results of these fits can be found in Tables~\ref{ParamEW}, \ref{Param} and \ref{Obs}. Note that the posteriors of EW fit parameters remain practically unchanged once NP effects via $\varepsilon_{ij}$ are included. It worth of mention that this is the case for $G_F$ (due to its precise measurement) which is used as a fit parameter, yet not for $G_F^{\mathcal L}$ whose posterior accidentally assumes (to the precision we are working at) the same value, $(1.16755 \pm 0.00089)\times 10^{-5}$, within scenario NP-I and NP-II.

This is, however, not the case for $V_{us}$ where the SM value of $0.2257(3)$ significantly changes once NP is included. Note that the main tensions within the SM originate from $Z\to\nu\nu$, $g_V^\ell$, and the conflicting determinations of $V_{us}$ from $K_L\to\pi\mu\nu$ and super-allowed beta decays. These are also the observables where the most significant improvements (compared to the SM scenario) are achieved in the NP case while the tension in $W\to\tau\nu/W\to e\nu$ cannot be resolved in our NP scenario.

The fit in the $\varepsilon_{ee}$-$\varepsilon_{\mu\mu}$ plane is depicted in the left plot of Fig.~\ref{FitReg} where, in addition, the preferred regions from the individual classes of processes are shown. The fact that all regions overlap at the $68\%$~CL reflects the goodness of the fit.

In order to better judge the agreement of the NP hypotheses with data and how this compares to the SM, we define the pull (with respect to the SM) for an observable $O_i$ as
\begin{align}
P(O_i) = \left|\frac{O_i^{\text{exp}}-O_i^{\text{SM}}}{\sqrt{(\sigma_i^{\text{exp}})^2+(\sigma_i^{\text{SM}})^2}}\right| - \left|\frac{O_i^{\text{exp}}-O_i^{\text{NP}}}{\sqrt{(\sigma_i^{\text{exp}})^2+(\sigma_i^{\text{NP}})^2}}\right|\,.
\end{align}
These pulls are reported in Table~\ref{Obs} for the two NP scenarios, where also the SM and NP posteriors for all observables (except the EW fit parameters shown in Fig.~\ref{ParamEW} which are unaffected by the NP parameters) are shown. Pulls signalling a better agreement between NP and data (with respect to the SM) are highlighted in gray. 

Finally, we show in the right plot of Fig.~\ref{FitReg} the constraints in the $\varepsilon_{ee}\varepsilon_{\mu\mu}$ plane for the case in which the modified gauge boson couplings to leptons are generated by a right-handed neutrino.

\bibliography{bibliography}

\flushcolsend

\end{document}